\documentclass[a4paper,11pt]{article}
\usepackage{graphicx,amssymb,amsmath,bm,latexsym,color,epsf,cite,comment}
\makeatletter
\@addtoreset{equation}{section}

\makeatother
\newcommand{\bea}{\begin{eqnarray}}
\newcommand{\eea}{\end{eqnarray}}
\newcommand{\vs}[1]{\vspace{#1 mm}}
\newcommand{\hs}[1]{\hspace{#1 mm}}
\renewcommand{\a}{\alpha}
\renewcommand{\b}{\beta}
\renewcommand{\d}{\delta}

\newcommand{\s}{\sigma}
\renewcommand{\t}{\theta}
\newcommand{\G}{\Gamma}

\newcommand{\la}{\lambda}
\newcommand{\pa}{\partial}
\newcommand{\nn}{\nonumber\\}

\newcommand{\Tr}{{\rm Tr}}

\newcommand{\tL}{{\tilde\Lambda}}
\newcommand{\tG}{{\tilde G}}

\begin{document}

%\begin{flushright}
% \\
%\today
%\end{flushright}

\begin{center}
{\Large\bf Wave Function Renormalization and Flow of Couplings\\
\vs{3}
 in Asymptotically Safe Quantum Gravity}
\vs{10}

{\large
Hikaru Kawai$^{a,b,}$\footnote{e-mail address: hikarukawai@phys.ntu.edu.tw}
and
Nobuyoshi Ohta$^{c,d,}$\footnote{e-mail address: ohtan@ncu.edu.tw}
} \\
\vs{5}

$^a${\em Department of Physics and Center for Theoretical Physics, National Taiwan University,
Taipei 106, Taiwan}

$^b${\em Physics Division, National Center for Theoretical Sciences,
Taipei 106, Taiwan}
\vs{2}

$^c${\em Department of Physics, National Central University, Zhongli, Taoyuan 320317, Taiwan}

and

$^d${\em Research Institute for Science and Technology, Kindai University,\\
Higashi-Osaka, Osaka 577-8502, Japan}

\vs{10}
%%%%%%%%%%%%%%%%%%%%%%%%%%%%%%%%
{\bf Abstract}
\end{center}

The importance of the proper treatment of the wave function renormalization in the renormalization group analysis
of quantum gravity is pointed out. The renormalization factor, sometimes called an inessential coupling, can
be used to fix any one of the coupling constants, with the exception of the coupling constants that remain
unchanged by the rescaling of the field.
Choosing to fix the cosmological constant, we propose to use a new regulator to obtain the renormalization
group equations for invariant couplings which tell us the flow of the Newton and $R^2$ couplings.
We find that the Newton coupling reaches a nontrivial ultraviolet fixed point (FP) and becomes small
in the low energy, but find only asymptotically free FP of the $R^2$ couplings for the higher-derivative
gauge fixing and regulator.
For the asymptotically free FP, we find that both of
the two independent terms are relevant operators in the high energy.
It is noted that the existence of nontrivial FPs may depend on the choice of the gauge and regulator.

\setcounter{footnote}{0}

%%%%%%%%%%%%%%%%%%%%%%%%%%%%
\section{Introduction}
%%%%%%%%%%%%%%%%%%%%%%%%%%%%

One of the longstanding problems in particle physics and gravity is to formulate the quantum theory of gravity
in a consistent manner. String theory is a promising candidate for such a theory.
However, in order to make physically meaningful predictions, nonperturbative effects must be incorporated.
So far, however, attempts to do so have not been entirely successful.

On the other hand, it seems that the possibility of formulating quantum gravity in the context of
ordinary field theory is still open.
After the idea of the asymptotic safety was initiated in~\cite{Weinberg}, many considerations have been made
along this line. The main problem here is how to describe the renormalization group flow for
a nonrenormalizable theory like gravity. One possibility is to use the ordinary perturbation theory
near two dimensions and apply the $\epsilon$-expansion as was proposed in~\cite{Weinberg}.
It turns out that the pole structure of the $2+\epsilon$ dimensional gravity is rather complicated than
was expected but still $\epsilon$-expansion is possible~\cite{KN}. However it is hard to tell whether
a fixed point (FP) exists in four dimensions or not, because four is too far from two.
Another possibility is to try to find a FP numerically based on constructive definition of gravity.
In fact several works have been made in this direction, and there appeared some results that support
the existence of a FP~\cite{D1,D2,D3}.
The most important question here is whether such FPs correspond to unitary field theories.
The problem stems from the fact that unitarity is not guaranteed in standard dynamical triangulation.
When unitarity is not guaranteed, the simplest class of FPs would correspond to $R^2$-type theories with ghosts.
Therefore, when finding a FP, one must be sure that it is a different theory from
the simple $R^2$-type theory. One exception is the causal dynamical triangulation in which
unitarity is manifest~\cite{AL,AGJL}.
In this case instead, we should check whether Lorentz invariance is recovered or not.
In other words, one needs to show that the FP is sufficiently far from the perturbative theory of
the  $R^2$ action. This was the situation around the mid-1990s, and no definite answers have been
made to this question of unitarity.

There have been several active attempts~\cite{Reuter,NR,perbook,rsbook}, based on the nonperturbative functional
renormalization group equation (FRGE)~\cite{Wetterich,Morris} to seek for asymptotically safe unitary theory.
The explicit form of the FRGE is given in terms of the  effective average action $\G_k$ as
\bea
\dot \G_k=\frac12 \Tr \left(\G_k^{(2)}+R_k\right)^{-1}\dot R_k,
\label{flow}
\eea
where $k$ is the cutoff, $R_k$ is a cutoff function suppressing the contribution of the modes below
the momentum scale $k$ to $\G_k$, the dot is a derivative with respect to $t=\log (k/k_0)$, $k_0$ being
an arbitrary initial value, and $\G_k^{(2)}$ is the second variation of the effective average action.
Although the flow equation~\eqref{flow} is exact, we have to make approximations to solve \eqref{flow}.
In general, the effective average action $\G_k$ includes an infinite number of all possible general coordinate
invariant operators, called theory space, generated by quantum effects.
The FRGE ~\eqref{flow}
gives flows in this infinite-dimensional theory space.
However, in order to solve the flow equation~\eqref{flow}, we have to make approximation by restricting
this infinite-dimensional theory space to a finite subspace. This approximation is called truncation.
The approach of asymptotic safety attempts to find nontrivial FPs away from asymptotically free theory
for a given truncation.
Once such fixed points are found, we have to extend the restricted space to study whether additional operators
are necessary to define the ultraviolet (UV) theory until there are no more necessary operators.

In general, any field theory contains redundant operators whose  coupling parameters change when we subject
the fields to a point transformation and have no effects on physical quantities.
It is well known that they need to be treated properly in order to study
FP actions. In quantum gravity, the couplings of such redundant operators are sometimes referred to as
inessential couplings~\cite{Weinberg}. The most typical of these is the wave function renormalization.
This appears in any truncation of the action because it preserves the form of each term.
In ordinary field theory, we usually fix the coefficient of the kinetic term by the wave function renormalization.
Any one of the coupling constants, however, can be fixed, except those that are unchanged by
the field rescaling.
In the case of quantum gravity, which consists of the Einstein, cosmological constant and $R^2$ terms, either
the Newton coupling or the cosmological constant can be fixed using wave function renormalization.
It does not make physical sense to consider the flow of these couplings separately.
The choice of a parameter to be fixed is not important because the physics depends only on the combination
of the coupling constants which does not depend on the wave function renormalization.
For example, if the cosmological constant is fixed, the renormalization group equation tells us the flow of
the Newton and $R^2$ couplings.
In the perturbative approach, this point was discussed in Ref.~\cite{FT}.
However until now, much attention has not been paid to this point in the context of asymptotically safe gravity.
Here we take this point into account and study the flow of the coupling constants.
In this process, we find that the most often adopted cutoff function breaks the invariance under
the wave function renormalization. We propose to use a different cutoff function
to resolve the problem.
In this manner we find that the Newton coupling reaches nontrivial UV FP and becomes small
in the low energy, and that the $R^2$ couplings are asymptotically free.

Recently such field redefinition is also considered in \cite{BF,Knorr}.
The authors consider the redefinition of the metric not just by overall factor but also by Ricci tensor etc.
and neglect higher order terms that would appear by such a redefinition. This would completely change the truncated
theory, or such redefinition does not close at the level of quadratic curvature gravity.
Again in the perturbative approach to the quantum Einstein gravity, this kind of field redefinition
was considered long time ago, see for example \cite{TV,Solodukhin}.
Related discussions of parametrization of the metric were made in \cite{GKL,OPP1,OPP2,GPS,MRS}.
Here we only consider the effect of the wave function renormalization factor and
find rather nontrivial results.

In the asymptotic safety, operators whose couplings have UV FPs are called relevant operators,
while those whose couplings do not have UV FPs and diverge in UV are called irrelevant operators.
The formers are the operators that should be kept in order to get well-defined UV behavior,
while the latter should be adjusted to be absent in the theory.
For example, in perturbation theory (corresponding to trivial FP), operators with dimensions less than
or equal to four are relevant operators and we have to keep these to obtain perturbatively renormalizable theory.
The expectation of asymptotic safety is that the number of relevant operators needed to define a theory
is finite, allowing one to make predictions about physical quantities.
Such a theory is called nonperturbatively renormalizable theory.

Much evidence has accumulated to show that such FPs exist, see for example~\cite{CPR,BMS,NR,perbook,rsbook}
and references therein.
An important problem in the asymptotically safe theory is to find how many relevant operators we have
and to identify these operators.
In perturbation theory, it has been known for some time that gravity with quadratic curvature terms are
renormalizable~\cite{Stelle}. There are two kinds of independent diffeomorphism invariant operators
in four dimensions, which can be chosen Weyl curvature squared and scalar curvature squared.
So we consider the action
\bea
S = \int  d^4 x \sqrt{g}\, \left[2\Lambda- Z_N R +\frac{1}{2\la}C_{\mu\nu\rho\s}{}^2+\frac{1}{\xi} R^2
-\frac{1}{\rho}E \right],
\label{bareaction}
\eea
where $Z_N=\frac{1}{16\pi G_N}$ with the Newton coupling $G_N$, $\Lambda$ is the cosmological constant,
$C_{\mu\nu\rho\s}$ the Weyl tensor, $R$ the scalar curvature and $E$ is the Gauss-Bonnet term which
is a total derivative and topological in four dimensions. As such, $\rho$ does not contribute
to the quantum effects so we may neglect this term in our study.
Note that the definition of the cosmological constant is different from, say Ref.~\cite{CPR},
and $\la,\, \xi$ and $\rho$ are dimensionless couplings.
These operators are necessary in order for the theory to be renormalizable.
This suggests that, together with the cosmological constant and the Newton coupling, there are
four relevant operators.

As mentioned above, operators with dimensions less than or equal to four are relevant for asymptotically free FP.
However this is a nontrivial problem in the asymptotically safe theory, because if the couplings
at the UV FP are nonzero, it is possible that quantum effects may change their properties.
The beta functions of these couplings have been calculated
in the literature~\cite{AB,BPS,CP,CPR,Niedermaier,GRSZ,OP1,OP2,Ohta:2022}, and it was thought
that there was only asymptotically free FP.
It turned out that if we keep contributions from the Newton coupling, there are nontrivial FP~\cite{BMS,FLNR,FOP}.
A surprise is that it has been claimed that there are only three relevant operators at the nontrivial FPs
of $\la$ and $\xi$~\cite{BMS,FLNR}.
However the analysis was not complete because the calculation was made on the Einstein space. Then the two
quadratic curvature terms are not independent in four dimensions, and these beta functions cannot be uniquely
determined.
In Ref.~\cite{FOP}, the beta functions are calculated on the general backgrounds, and it was found that
there are indeed only three relevant operators.
Unfortunately the calculation was made in the expansion in the inverse Newton couping, and then
it is not possible to see if the nontrivial FPs found there may be smoothly connected to the low energy,
where we expect that the Newton coupling becomes small.
This is important in studying how the quantum effects affect physical quantities at various energies,
in particular in the context of cosmology.
In order to study this, we have to calculate the beta functions without expanding in the inverse Newton coupling.
Here we present the result including all orders in the inverse Newton coupling and study the flow of couplings
between the low and high energies with the higher-derivative gauge fixing and regulator.

Unfortunately we have not been able to find any nontrivial FP in the range of $G_N>0$ and $\la>0$,
though we have searched for it in the wide range of parameter space. Indeed, one of the present authors (N.O.)
together with K. Falls and R. Percacci searched for such nontrivial FPs in the higher-derivative
couplings with higher-derivative regulators without taking into account the wave function renormalization,
and did not find any reasonable nontrivial FP.
On the other hand, some nontrivial FPs were found if the standard Feynman-De Donder gauge and
lower derivative regulator was used~\cite{FOP2}.%
\footnote{ Similar FPs were found in \cite{SWY}.}
It was not clear if such FPs, depending on the gauge and/or regulator, are physical or not.
So here we concentrate on the analysis of the flow of the asymptotically free FP, which is universal.

This paper is organized as follows. In sect.~\ref{Ein}, we start with the analysis of the Einstein theory
with the cosmological constant term.
In subsect.~\ref{earlier}, we recapitulate the results of the FPs when we do not consider the freedom
of wave  function renormalization.
In subsect.~\ref{inessential}, we consider the effect of wave function renormalization and note that
the cosmological constant and Newton coupling change by the wave function renormalization.
Then we formulate the FRGE with such a freedom, and use this freedom to fix the cosmological constant.
In this process, we show that we should use a new cutoff which does not break the invariance under
the wave function renormalization. Using this regularization, we find the beta functions are
entirely written in terms of the coupling constants invariant under the wave function renormalization.
We find that the Newton coupling go to UV FP and becomes smaller in the low energy, as expected.
In sect.~\ref{quadratic}, we discuss the quadratic curvature theory. Here we have not found physically
reasonable nontrivial FPs for the dimensionless couplings $\la$ and $\xi$, but have found
asymptotically free FP with the higher-derivative gauge fixing term and regulator.
For the asymptotically free FP, we show that both the Weyl curvature square and
scalar curvature square are relevant operators, in agreement with the perturbation theory.
Again the Newton coupling is found to go to UV FP and becomes smaller in the low energy.
Section~\ref{conclusion} is devoted to summary and discussions.

\section{Einstein gravity with the cosmological constant}
\label{Ein}

As a warm-up, let us study the Einstein theory with cosmological constant.
That is, we will consider only the cosmological constant and the Einstein term in \eqref{bareaction},
and look at the results when the wave function renormalization is properly handled.

\subsection{Earlier results}
\label{earlier}

First, we check the case where the wave function renormalization is not considered.
Using the type Ia cutoff (in the standard Feynman-De Donder gauge)
\bea
R(\Delta)= Z_N (k^2-\Delta) \t(k^2-\Delta),
\label{old_cutoff}
\eea
where $\Delta \equiv - g^{\mu\nu} \nabla_\mu\nabla_\nu$, we get the FRGE~\cite{CPR}
\bea
2(\dot{\tL}+4\tL)=\frac{1}{16\pi}(A_1+A_2\eta_Z), \nn
\frac{\dot{\tG}-2 \tG}{16\pi \tG^2}=\frac{1}{16\pi}(B_1+B_2\eta_Z),
\label{frge_ein}
\eea
where the dimensionless couplings are defined by
\bea
\Lambda = k^4 \tL, \qquad
G_N = k^{-2}\tG,
\label{scale}
\eea
and the quantities on the right-hand side (rhs) are given by
\bea
&& A_1 = \frac{1+128\pi \tG \tL}{\pi(1-32\pi \tG \tL)},\qquad
A_2 = \frac{5}{6\pi(1-32\pi \tG \tL)},\nn
&& B_1 = - \frac{11-288\pi \tG \tL+7 (32 \pi \tG \tL)^2}{3\pi(1-32\pi \tG \tL)^2},\quad
B_2 = -\frac{1+160 \pi \tG \tL}{12\pi(1-32\pi \tG \tL)^2},
\label{coef_old}
\eea
which are obtained from those in \cite{CPR} by the replacement $\tL\to 16\pi \tL\tG$.
The anomalous dimension $\eta_Z$ for the inverse of the Newton coupling is defined by
\bea
\eta_Z = -\frac{\dot Z_N}{Z_N}=2-\frac{\dot\tG}{\tG}.
\eea

The beta functions for the cosmological constant and the Newton coupling are obtained from eqs.~\eqref{frge_ein} as
\bea
\dot{\tilde\Lambda} &=& -4\tilde\Lambda +\frac{A_1+(A_1 B_2-A_2 B_1)\tilde G}{32 \pi(1+B_2 \tilde G)}, \nn
\dot{\tilde G} &=& 2\tilde G +\frac{B_1 {\tilde G}^2}{1+B_2 \tilde G}.
\label{betas_old}
\eea
We find that the FP for positive Newton coupling is located at
\bea
(\tL_*,{\tG}{}_*)=(0.00543, 0.7073).
\label{fp1}
\eea
This result is consistent with \cite{CPR} because the cosmological constant there
is obtained by $16\pi \tL_* {\tG}{}_* =0.193$.
In addition, there is another FP at $(\tL_*,{\tG}{}_*)=(0.0007916, 0)$,
which should correspond to the Gaussian FP in the usual convention.

\subsection{Inessential coupling: The wave function renormalization}
\label{inessential}

Next, we examine the effect of the wave function renormalization.
If we express the metric in terms of the rescaled (or renormalized) metric as
\bea
g_{\mu\nu} = Z g'_{\mu\nu},
\label{sc}
\eea
the terms in the action~\eqref{bareaction} can be written as
\bea
\sqrt{g} = Z^2 \sqrt{g'}, \qquad
\sqrt{g} R = Z \sqrt{g'} R',
\eea
and the other terms are unchanged.
This means that under the wave function renormalization, the couplings scale as
\bea
\Lambda'= Z^{2} \Lambda, \qquad
G_N' = Z^{-1} G_N.
\eea
In other words, the following identity holds
\bea
S_{eff}[Z g'_{\mu\nu}, \Lambda(t), G_N(t), t]
= S_{eff}[g'_{\mu\nu},Z^2 \Lambda(t), Z^{-1} G_N(t), t].
\eea
provided that the energy scale parameter $t$ does not change under the rescaling (\ref{sc}).
Note that the dimensionless combination $\Lambda G_N^2$ is invariant under this rescaling.

By definition, the effective action in terms of $g'_{\mu\nu}$ is given by
\bea
S'_{eff}[g'_{\mu\nu},\Lambda(t), G_N(t), t]= S_{eff}[ g_{\mu\nu},\Lambda(t), G_N(t), t],
\eea
and it is evaluated as
\bea
= S_{eff}[Z g'_{\mu\nu}, \Lambda(t), G_N(t), t]
= S_{eff}[g'_{\mu\nu},Z^{2} \Lambda(t), Z^{-1} G_N(t), t].
\label{inv}
\eea
Thus, if we introduce a wave function renormalization at each step of the renormalization transformation,
it will generate additional infinitesimal transformations as
\bea
\d\Lambda=2\, \zeta \Lambda, \qquad
\d G_N =- \zeta G_N.
\eea
The FRGE~\eqref{frge_ein} is modified as
\bea
\dot{\tL}+4\tL=\frac{1}{32\pi}(A_1+A_2\eta_Z)+2 \zeta \tL, \nn
\dot{\tG}-2 \tG=\tG^2 (B_1+B_2\eta_Z) - \zeta \tG,
\label{betas_new}
\eea
Because of the freedom of wave function renormalization, we should note that the values of $\tL$
or $\tG$ themselves do not have physical meaning separately. Only the combination of $\tL \tG^2$ has
physical meaning. Using this freedom, we can impose the condition that either the (dimensionless) cosmological constant
or the Newton coupling is fixed for the whole range of $t$, or consider only the FRGE for
the invariant combination $\tL \tG^2$.

We must remember that the above analysis is based on the assumption that the energy scale $k$ (or $t$) is
invariant under wave function renormalization. It can be seen that the cutoff~\eqref{old_cutoff} commonly
used in the literature clearly violates this condition. This is because under the rescaling of the metric
$\Delta (\equiv -g^{\mu\nu}\nabla_\mu \nabla_\nu)$ is transformed to $Z^{-1} \Delta$.
To correct this, we propose to use the following cutoff:
\bea
R(\Delta)= Z_N (\sqrt{\tL}\, k^2-\Delta) \t(\sqrt{\tL}\, k^2-\Delta).
\label{new_cutoff}
\eea
Since we are going to fix $\tilde\Lambda$, this in effect introduces a numerical factor to $k^2$ in
\eqref{old_cutoff}, but the factor is important to make the term transform consistently under the rescaling.
Using the cutoff~\eqref{new_cutoff}, we find that the quantities in \eqref{coef_old} are modified to
\bea
A_1 &=& \frac{(\dot{\tL}+4\tL)(1+128\pi \tG \sqrt{\tL}\,\big)}{4\pi(1-32\pi \tG \sqrt{\tL})}, \nn
A_2 &=& \frac{5\tL}{6\pi(1-32\pi \tG \sqrt{\tL})},\nn
B_1 &=& - \frac{(\dot{\tL}+4\tL)\Big[11-288\pi \tG \sqrt{\tL}+7 (32 \pi \tG \sqrt{\tL}\,\big)^2\Big]}
{12\pi(1-32\pi \tG \sqrt{\tL})^2 \sqrt{\tL}} , \nn
B_2 &=& -\frac{(1+160 \pi \tG \sqrt{\tL})\sqrt{\tL}}{12\pi(1-32\pi \tG \sqrt{\tL}\,\big)^2}.\hs{10}
\label{coef_new}
\eea
Note that here we still have $\dot\tL$ in $A_1$ and $B_1$ on the rhs.
Imposing that the $\tL$ is constant, we can obtain $\zeta$ from \eqref{betas_new}.
Substituting this into the second equation in~\eqref{betas_new}, we find
\bea
\dot\tG=\frac{2(8-19\eta+\eta^2-14\eta^3)\tG}{5-6\eta-5\eta^2+384\pi^2(1-\eta)^2},
\label{FRGEpre}
\eea
where we have defined the invariant variable
\bea
\eta=32 \pi \tG\sqrt{\tL}.
\eea
Equation~\eqref{FRGEpre} can be rewritten into the FRGE in terms of $\eta$:
\bea
\dot \eta = \frac{2(8-19\eta+\eta^2-14\eta^3)\eta}{5-6\eta-5\eta^2+384\pi^2(1-\eta)^2}.
\eea
It is important to notice that the rhs is solely written in terms of $\eta$, the invariant combination.
If we did not use the modified cutoff~\eqref{new_cutoff}, it would depend on other combination and
the flow equation is inconsistent. For example, if we used the cutoff~\eqref{old_cutoff}, we find that
the rhs is a function of $\tL\tG$. However this is not invariant under the wave function renormalization.

\begin{figure}[tb]
\begin{center}
\includegraphics[width=70mm]{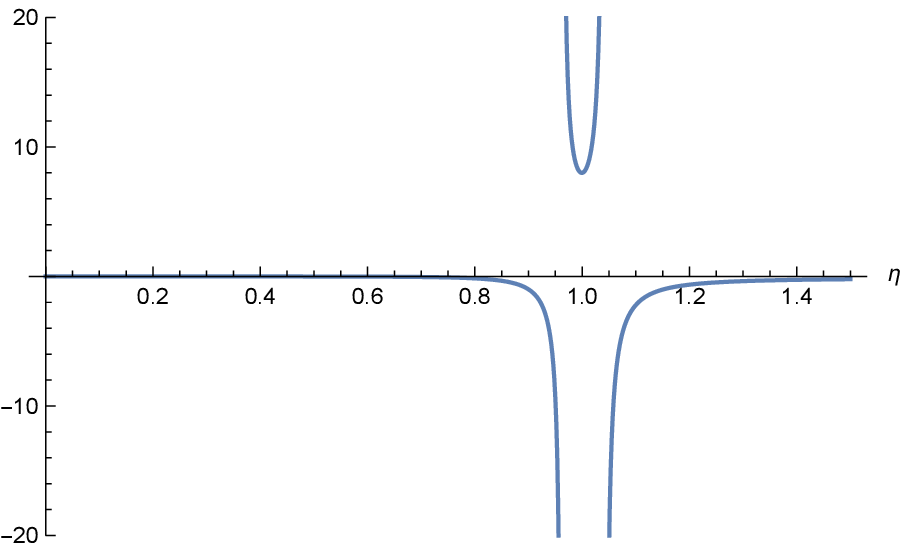}
\put(-100,-20){(a)}
\hs{10}
\includegraphics[width=70mm]{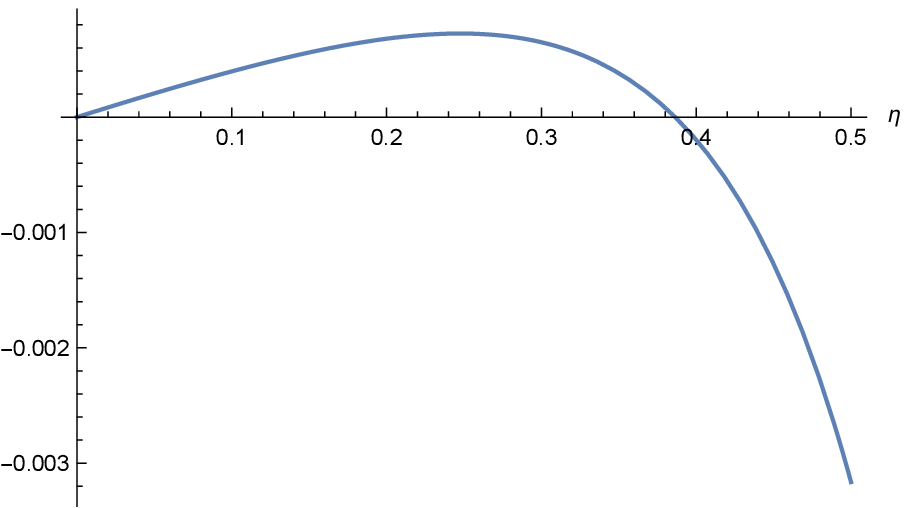}
\put(-100,-20){(b)}
\end{center}
\caption{The beta function of $\eta$ for the range of (a) $0<\eta<1.5$ and (b) $0<\eta<0.5$.}
\label{f1}
\end{figure}

We plot the beta function for $\eta$ in Fig.~\ref{f1}.
This is a typical asymptotically safe theory for $0 \leq \eta \leq 0.3864$.
The beta function has only one positive zero:
\bea
\eta_* = 0.3864,
\eea
which corresponds to the UV FP.

\begin{figure}[tbh]
\begin{center}
\includegraphics[width=70mm]{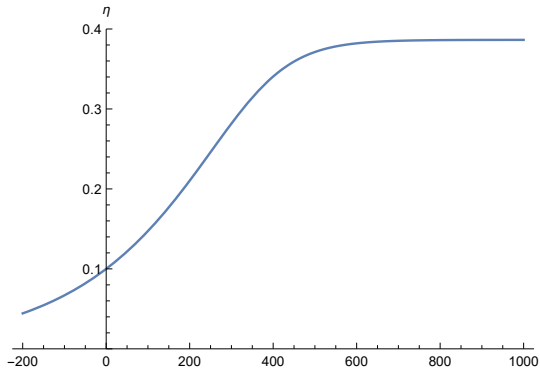}
\end{center}
\caption{The flow of $\eta$ for the boundary condition $\eta=0.1$ at $t=0$.}
\label{f2}
\end{figure}

We expect the usual dimensionful Newton coupling, which describes our world,
remains small but finite in the low energy. We see from \eqref{scale} that this means that
the dimensionless Newton coupling $\tG$ becomes small in the low energy.
We plot the flow of $\eta$ from the low energy ($t=-200$) to the high energy ($t=1000$)
for the boundary condition $\eta=0.1$ at $t=0$.
We can fix the cosmological constant $\tL$ to a certain value, and in this case, the FRGE gives the flow of
the Newton coupling. We see that the Newton coupling goes to the FP at the high energy and flows
down to small value in the low energy, as shown in Fig.~\ref{f2}. This is the expected behavior.

The stability matrices are
\bea
\left. \frac{\pa \b_\eta}{\pa \eta}\right|_{\eta_*}=-0.01325,
\label{fp2}
\eea
This means that this FP is UV stable. This is actually obvious from the behavior of $\b_\eta$
shown in Fig.~\ref{f1} (b).

%\newpage
\section{Quadratic gravity}
\label{quadratic}

In Ref.~\cite{FOP}, the beta functions for the quadratic gravity~\eqref{bareaction}  on general backgrounds
were calculated to the first order in $Z_N$. Then it is found that there are nontrivial FPs for the beta functions
for $\la$ and $\xi$, and it is discussed how many relevant operators exist.
It is found that only three out of the couplings $\tL, \tG, \la$ and $\xi$ are relevant,
in sharp contrast to the perturbation theory in which we need all four couplings for the renormalizability.
In the case of asymptotically free FP $\la_*=\xi_*=0$, the ratio of these couplings is finite with
$\omega_* \equiv  -\frac{3\la_*}{\xi_*}=-0.02286$.

However, only the first order of the inverse of the Newton coupling $\tG$ is kept in the above approximation,
and we cannot study how the FPs are connected to the low energy where we expect that the Newton
coupling becomes small. In order to study this and settle the question of the number of relevant operators,
we have to calculate the beta functions for all orders in $Z_N$.
However this calculation is quite involved. Fortunately Benjamin Knorr developed a code which enabled us
to calculate the beta functions.

The calculation done in \cite{FOP} used the higher derivative cutoff
\bea
R(\Delta) \sim (k^4-\Delta^2) \t(k^4-\Delta^2),
\eea
where the front factor, which is chosen to match the higher derivative term in the Hessian, is not exposed explicitly.
However, as pointed out in the Einstein theory, we have to modify this to be consistent with the scaling
by the wave function renormalization. Here we take
\bea
R(\Delta) \sim (\tL k^4-\Delta^2)\, \t(\tL k^4-\Delta^2).
\label{high_cutoff}
\eea
By using Knorr's program, we have calculated the flow equations:
\bea
\dot{\tL} &=&  - 4 \tL + \tL\, f_1(\eta,\la,\xi) +2\zeta \tL, \nn
\dot\tG &=& 2\tG + 16 \pi \tG^2 \sqrt{\tL}\, f_2(\eta,\la,\xi)-\zeta \tG, \nn
\dot\la &=& -2 \la^2 f_3(\eta,\la,\xi), \nn
\dot \xi &=& -\xi^2 f_4(\eta,\la,\xi).
\eea
The explicit form of the functions $f_1, \cdots, f_4$ are too complicated to be presented here.
Here we have already incorporated the effects by the wave function renormalization in the equations for
$\tL$ and $\tG$. Note that $\la$ and $\xi$ are dimensionless couplings, and are not affected by
the wave function renormalization.
So their flow equations take already invariant form under the transformation of wave function renormalization
and involve only the invariant combination $\eta$.

As before, we impose the condition to fix the cosmological constant to obtain
\bea
\zeta = 2 -\frac12 f_1(\eta,\la,\xi).
\eea
Substituting this into the beta function for $\tG$, we get
\bea
\dot\tG = \frac12 \tG f_1(\eta,\la,\xi) +16\pi\tG^2 \sqrt{\tL} f_2(\eta,\la,\xi).
\eea
This can be cast into the equation for $\eta$:
\bea
\dot \eta = \frac{\eta}{2} f_1(\eta,\la,\xi)+\frac{\eta^2}{2} f_2(\eta,\la,\xi).
\eea
We thus find the flow equation invariant under the wave function renormalization.
The explicit forms of the beta functions for $\eta, \la$ and $\xi$ are given in the Mathematica file as ancillary file
in the arXiv page.

We have found one nontrivial FP
\bea
\eta_* = 2.1397, \qquad
\la_* = -1.3731, \qquad
\xi_* = -7.7864,
\label{ntfp}
\eea
but this is probably unphysical FP because $\la$ is supposed to be positive for the stability of the system.
A similar FP is found in~\cite{SWY} though they have not taken into account the fact
that the wave function renormalization is inessential.
Unfortunately we have not been able to find any other reasonable FP except for the asymptotically free FP.
As mentioned in the Introduction, this may be an artifact of our
gauge and/or regulator, or our search may not be enough.
Whichever it is, the existence of the asymptotically free FP is universal.
So here we study the flow of the asymptotically free (but nonzero FP for $\tG$) FP.
When we study asymptotically free FP, it is more convenient to use $\omega=\frac{-3\la}{\xi}$
rather than $\xi$. We find the FP
\bea
\eta_* = 0.7788, \qquad
\la_* = 0, \qquad
\omega_* = -0.02286,
\label{asfp}
\eea
In this case, both $\la$ and $\xi$ go to zero asymptotically with their ratio $\omega$ being fixed.

In order to see whether this FP is relevant or not, we make expansion to the second order of
the beta functions for small $\la$ and $\xi$. We find
\bea
\b_\la &=& -\frac{3591+796\xi}{480\pi^2(9+2\xi)}\la^2 +O(\la^3)= -\frac{133}{160\pi^2}\la^2 +O(\la^3,\la^2\xi), \nn
\b_\xi &=& -\frac{5\xi^2}{576\pi^2}+\left(\frac{5\xi}{16\pi^2}+\frac{(163-700\eta)\xi^2}{1080\pi^2 \eta}\right)\la \nn
&& -\left(\frac{5}{8\pi^2}+\frac{(44-145\eta)\xi}{72\pi^2 \eta}+\frac{(6385-9288\eta+21800\eta^2)\xi^2}{6480\pi^2
 \eta^2}\right)\la^2
+O(\xi^3,\xi^2\la,\la^3)\nn
&=& -\frac{5(72\la^2-36\la\xi+\xi^2)}{576\pi^2} + O(\xi^3,\xi^2\la,\la^3),
\label{R2b}
\eea
in agreement with the known results~\cite{FT,AB,BPS,CP,CPR,Niedermaier,GRSZ,OP1,OP2,Ohta:2022,FOP}.
%Here we assume that $\eta$ is finite.
This gives a nontrivial check of our results.

If we simply check the stability matrix (defined by the first derivatives of the beta functions at the FP),
its eigenvalues are zero. So naively these are marginal operators. In this case, we should look at
the higher order terms in the couplings.
It is clear from the beta function for $\la$, $\la$ is actually a relevant coupling near the origin.
Let us study the behavior of these coupling near the origin in more detail.

\begin{comment}
What about $\xi$? Its beta function is quadratic form in the couplings $\la$ and $\xi$, and
whether $\xi$ is relevant or not depends on which direction it approaches to zero.
If we substitute $\la=-\omega \xi/3$ into the beta function of $\xi$, we get
\bea
\b_\xi=-\frac{5(8\omega^2+12\omega+1)}{576\pi^2} \xi^2.
\eea
Thus we find that if $(8\omega^2+12\omega+1)>0$ or
\bea
\omega< -1.4114 \mbox{ or } -0.0886<\omega,
\eea
$\xi$ is a relevant coupling, otherwise it is irrelevant.
We then see from eq.~\eqref{asfp} that $\xi$ is asymptotically free and relevant.
This result is in agreement with perturbative treatment, where all couplings are necessary for the renormalizability.
\end{comment}

As is well known, the perturbative regions are asymptotically free or not, depending on the initial values
of the couplings constants. To see this, let us consider the renormalization group trajectory given by
\bea
\frac{d\xi}{d\lambda}= \frac{\beta_\xi}{\beta_\lambda}=a+b\frac{\xi}{\lambda}+c\frac{\xi^2}{\lambda^2},
\label{rt}
\eea
where we have dropped higher order terms for small couplings.
Set
\bea
\chi=\frac{\xi}{\la},
\eea
and eq.~\eqref{rt} is rewritten as
\bea
\la \frac{d\chi}{d\la} &=& a+(b-1)\chi+c\chi^2 \nn
&\equiv& c(\chi-\alpha)(\chi-\beta).
\label{rch}
\eea
Let us consider the case
\bea
a>0,\qquad b<0,  \qquad c>0, \qquad (b-1)^2-4a c>0,
\eea
so that both $\alpha$ and $\beta$ are real and positive.\footnote{Their values corresponding to  (\ref{R2b}) are
$a=\frac{100}{133}, b=-\frac{50}{133}, c=\frac{25}{2394}$, and $\alpha=131.2, \ \beta=0.5487$.
They are related to $\omega$; in fact we have $\omega_{*}=-\frac{3}{\alpha}$.}
Without loss of generality, we can assume
\bea
\alpha>\beta.
\eea

\begin{figure}[tb]
\begin{center}
\includegraphics[width=60mm]{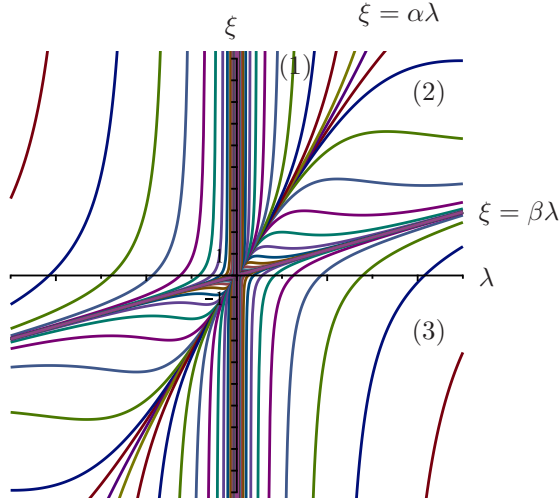}
\put(-40,180){\small $\xi=\a \la$}
\put(5,105){\small $\xi=\b \la$}
\put(-70,160){\small (1)}
\put(-20,150){\small (2)}
\put(-20,60){\small (3)}
\put(-90,175){\small $\xi$}
\put(5,80){\small $\la$}

\end{center}
\caption{Renormalization group flow. %The horizontal axis is $\lambda$ and the vertical axis is $\xi$.
The flow direction is from right to left. The slope of the steeper line ($\xi=\a\la$) is $\alpha$ and
that of the flatter line ($\xi=\b\la$) is $\beta$. All trajectories in the region between the flatter line
and the vertical axis are tangent to the steeper line near the origin,
and trajectories outside this region do not pass through the origin.}
\label{fR2}
\end{figure}

Using this, we can determine the region of initial values where
the renormalization group flow converges to the origin $\lambda=\xi=0.$
First, obviously the initial value of $\lambda$ should be positive, and the direction of flow should be
in the direction of decreasing $\lambda$. Then the half $\lambda$-$\xi$ plane ($\lambda>0$) splits into three sectors
(see Fig.~\ref{fR2}.):
\bea
(1)\ \xi>\alpha\la, \qquad (2)\ \alpha \la>\xi>\beta\la, \qquad (3)\ \beta\la>\xi.
\eea
From (\ref{rch}), it is clear that these three sectors are not mixed by the renormalization group flow.
More explicitly, the flow in each sector is described by a different branch of the solution to (\ref{rch}).
Actually, the trajectory of the second sector (2) is given by
\bea
\xi=(\sigma-\delta \tanh(c\delta\log|\lambda|+\kappa))\lambda,
\label{rt1}
\eea
where
\bea
\delta=\frac{1}{2}(\alpha-\beta), \qquad \sigma=\frac{1}{2}(\alpha+\beta),
\eea
and $\kappa$ is an integration constant.
On the other hand, the trajectory in the first (1) or the third sector (3) is given by
\bea
\xi=(\sigma-\delta \coth(c\delta\log|\lambda|+\kappa))\lambda.
\label{rt2}
\eea
depending on whether the argument of $\coth$ is negative or positive.

In the first sector (1), the trajectory corresponds to the part of (\ref{rt2}) where
the argument of $\coth$ is negative where $\coth<-1$. Therefore as $\lambda$ is decreased to zero, the argument
of the $\coth$ remains negative and the value of $\chi$ (inside of the parentheses) converges to $\alpha$.
In the second sector (2), $\tanh$ in (\ref{rt1}) causes no singularity and $\chi$ again converges to $\alpha$.
Therefore in the first two sectors, the flow converges to the origin along the curve converging to $\xi=\a\la$,
as depicted in Fig.~\ref{fR2}. This precisely corresponds to the FP value for $\omega$ in \eqref{asfp}.
On the other hand in the third sector (3), the trajectory corresponds to the part of (\ref{rt2}) where
the argument of $\coth$ is positive. Then the argument of $\coth$ becomes zero at some smaller value of
$\lambda(>0)$, which means the flow goes away to $\xi=-\infty$.
Thus, the domain of the asymptotically free FP is given by
\bea
\xi> \beta\la, \qquad \lambda>0 .
\eea
Given this result, we may say that these two $R^2$ operators are marginally relevant operators.

We have thus found the flow of asymptotically free theory, in agreement with perturbative treatment,
where all couplings are necessary for the renormalizability.
However we should note that our analysis is not just perturbative treatment, but shows that these couplings
are necessary from the viewpoint of renormalization group for the theory to be renormalizable.
A typical example of nonrenormalizable theory in this sense, though perturbatively renormalizable, is QED,
where the coupling is not asymptotically free and the theory suffers from Landau singularity.

\begin{figure}[tbh]
\begin{center}
\includegraphics[width=60mm]{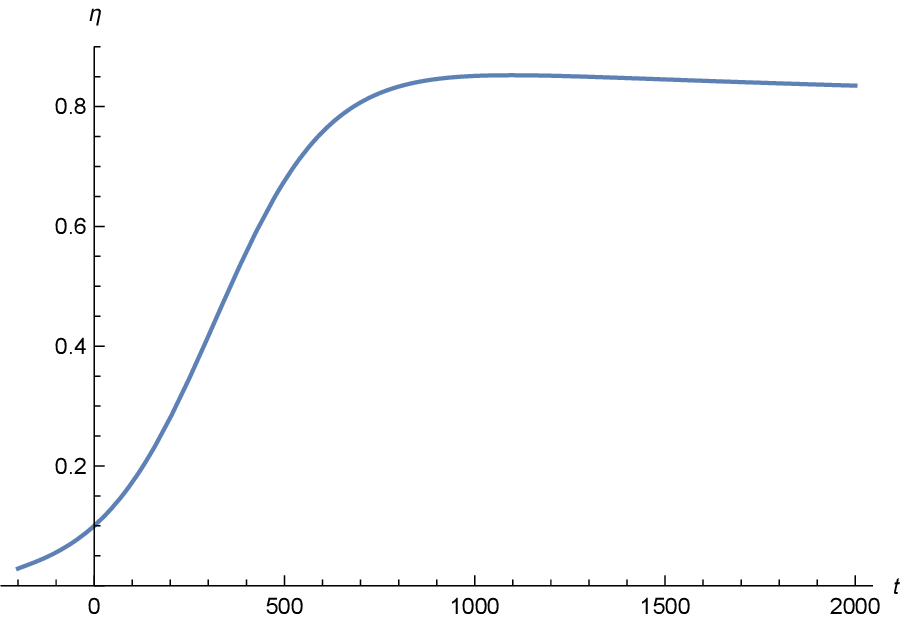}
\put(-80,-20){(a)}
\hs{20}
\includegraphics[width=60mm]{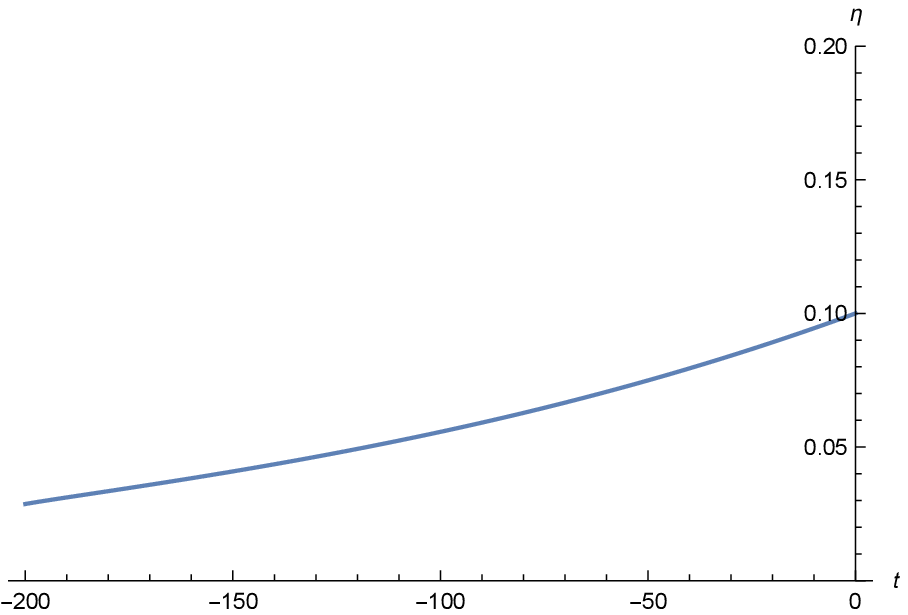}
\put(-80,-20){(b)}
%\end{center}
%\caption{The flow of $\eta$ for the range of (a) $-200<t<2000$ and $-200<t<0$.}
\\
\vs{5}
%\label{f3}
%\end{figure}
%\begin{figure}[h]
%\begin{center}
\includegraphics[width=60mm]{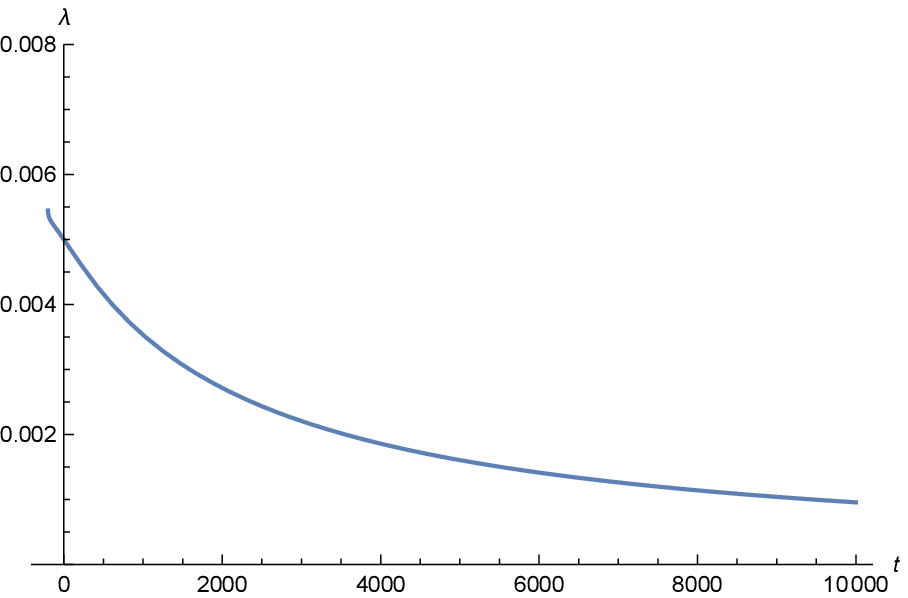}
\put(-80,-20){(c)}
\hs{20}
\includegraphics[width=60mm]{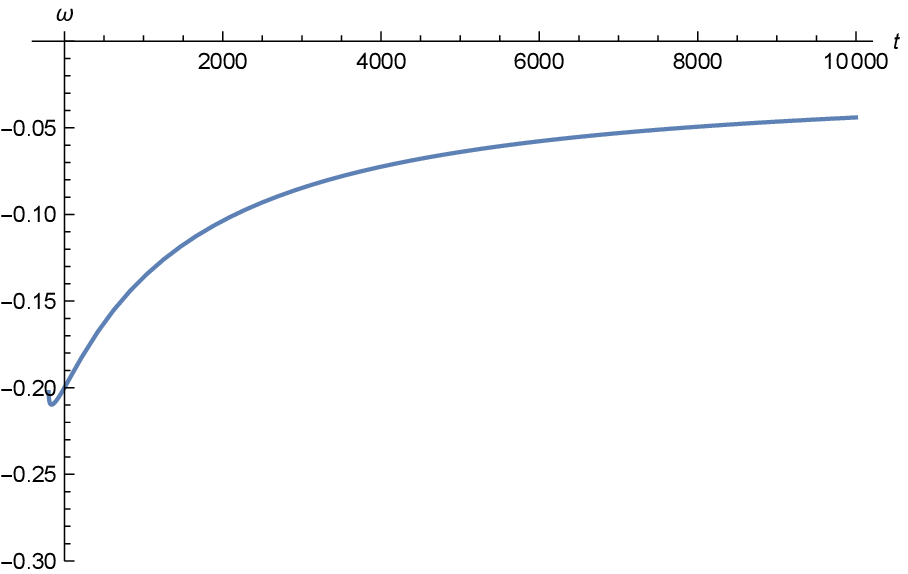}
\put(-80,-20){(d)}
\end{center}
\caption{The flow of $\eta$ for the range of (a) $-200<t<2000$ and (b) $-200<t<0$, and
the flows of (c) $\la$ and (d) $\omega$ for the range of $-200<t<10000$.}
\label{f4}
\end{figure}

We have also studied how the couplings flow from the low to high energies for the boundary condition
$\eta=0.1, \la=0.005,\omega=-0.2$ at $t=0$ in full order in the inverse Newton coupling without restricting
to the weak coupling expansion. The results are depicted in Figs.~\ref{f4}.
It is reassuring that we have precisely reproduced the expected behaviors for Newton coupling,
going to a finite UV FP and becoming smaller in IR.

\newpage
\section{Summary and discussions}
\label{conclusion}

We have studied the flow of the coupling constants in the Einstein and higher curvature theories.
In most of the discussions of FPs in the asymptotic safety, the effect of the wave function
renormalization have not been considered much, even though it was recognized from the outset to be inessential
coupling~\cite{Weinberg}. If we neglect this, it appears that we get FP both for the cosmological constant
and Newton coupling separately, but it is important to realize that they could be changed by the wave function
renormalization and do not make much sense.
The freedom in the choice of the wave function renormalization allows us to fix either the cosmological
constant or Newton coupling. Here we have chosen to fix the cosmological constant and studied the flow of
Newton and $R^2$ couplings. It should be noted that it is also perfectly right to fix the Newton coupling
to study the flow of the cosmological constant. Either way, the flow equation is written in terms of the invariant
combination $\eta$, and we find the same FP.

We have then studied the flow of the Newton coupling in the Einstein theory with the fixed cosmological constant.
In this process, we find that the widely accepted cutoff should be modified so as to be consistent with
the scaling by the wave function renormalization. In this way, we have found the beta functions are written
entirely in terms of invariant combination $\eta$ and found the nice behavior that the Newton coupling goes
to a finite UV FP and becomes small in the low energy, as expected.
We have also studied the behaviors of the couplings in quadratic curvature theory.
There we have found nontrivial FP for the cosmological constant and the Newton coupling with
asymptotically free FP for the higher curvature terms, in addition to nontrivial FP for the higher curvature
terms~\eqref{ntfp} which is probably unphysical. We find  here again that the beta functions are written
in terms of the invariant combination $\eta$.
We have given the detailed analysis of the asymptotically free FP for small couplings and revealed how the couplings go
to the free FP and identified the region where this happens around the origin of the coupling space.
They vanish in the UV with their ratio tending to a finite number.
We have also confirmed this behavior for the higher curvature terms and the expected behavior for
the Newton coupling using the full beta functions containing all order terms of the Newton coupling.
Our motivation to study the system is to examine how many relevant operators there are.
This is an important problem in identifying the asymptotically safe theory.
We have found that the two independent quadratic curvature terms are both marginally relevant
at the asymptotically free FP.
It is reassuring that we have been able to find a trajectory connecting between the high and low energies.

Unfortunately we have not been able to find any reasonable nontrivial UV FPs.
The importance of such FPs is that the asymptotically free theory is probably not able to resolve
the unitarity problem inherent in the higher-derivative theories. However, there are some suggestions that
the ghost may not be a problem because they may be confined in analogy with strong coupling QCD~\cite{HR},
or they are unstable massive modes which do not appear in the asymptotic states~\cite{DM}.
To determine whether either of these is true or not, or if there is another way to avoid the problem,
needs further confirmation.
So it is premature to conclude at this stage that the asymptotically free FP cannot save the theory.

We would also like to note that it is only that we have not been able to find good nontrivial UV FPs
with our present higher-derivative gauge fixing and regulator,
but this does not completely exclude the possibility that there are still sensible UV FPs somewhere
in the coupling space. This is because numerical calculations cannot rule out this possibility completely.
Another problem is that the result seems to depend on the choice of gauge and/or regulator~\cite{FOP2}.
To confirm the existence or nonexistence of the nontrivial FPs, we need further study.

Another possibility is that this failure is just an artifact of the truncation and it is possible to find
nontrivial FP if we extend our space of study to more operators.
This search for nontrivial FPs may end at certain stage without any more relevant and marginal operators,
or could continue.
The first case realizes the goal of the asymptotic safety program.
If the latter is true, there may be an infinite number of relevant operators.
In this case we have to understand how the coefficients of these terms are determined,
because otherwise the theory lacks its predictability. The principle to determine them may be
the requirement of the consistency of the theory, and then
the infinite series might be summed up to a closed theory,
just like massive gravity~\cite{dGT}.
This may lead to an alternative formulation of string theory or some other theory of quantum gravity
that can deal with phenomena in curved space.

\section*{Acknowledgments}

First we should thank Benjamin Knorr for calculating the full beta functions in the quadratic gravity
to all order in the Newton coupling and for valuable discussions.
We would also like to thank Kevin Falls and Roberto Percacci for numerous valuable discussions.
H.K. thanks Prof. Shin-Nan Yang and his family for their kind support through the Chin-Yu
chair professorship. H.K. is partially supported by JSPS (Grants-in-Aid for Scientific Research
Grants No. 20K03970), by the Ministry of Science and Technology, R.O.C.
(MOST 111-2811-M-002-016), and by National Taiwan University.
The work of N.O. was supported in part by the Grant-in-Aid for Scientific Research Fund of the JSPS (C) No. 16K05331,
No. 20K03980, and by the Ministry of Science and Technology, R. O. C. (Taiwan) under the grant MOST 111-2811-M-008-024.

\end{document}